\documentclass[11pt,twoside]{article}

\usepackage{asp2004}
\usepackage{epsf}
\usepackage{psfig}
\usepackage{lscape}

\begin{document}
\title{Infrared SEDs of Seyfert Galaxies: Starbursts and the Nature of the
Obscuring Medium} 

\author{Catherine L.\ Buchanan\altaffilmark{1}, Jack F.\
Gallimore\altaffilmark{2}, Christopher P.\ O'Dea\altaffilmark{1}, Stefi A.\
Baum\altaffilmark{3}, David J.\ Axon\altaffilmark{1}, Andrew
Robinson\altaffilmark{1}, Moshe Elitzur\altaffilmark{4}, \& Martin
Elvis\altaffilmark{5}}

\altaffiltext{1}{Department of Physics, Rochester Institute of
Technology, 54 Lomb Memorial Drive, Rochester NY 14623. Email:
clbsps@cis.rit.edu}
\altaffiltext{2}{Bucknell University Department of Physics, Moore Avenue,
Lewisburg, PA 17837}
\altaffiltext{3}{Center for Imaging Science, Rochester Institute of
Technology, 84 Lomb Memorial Drive, Rochester NY 14623. Email:
clbsps@cis.rit.edu}
\altaffiltext{4}{University of Kentucky Physics \& Astronomy Department,
Lexington, KY 40506}
\altaffiltext{5}{Harvard-Smithsonian Center for Astrophysics, 60 Garden St.,
Cambridge, MA 02138}

\begin{abstract}
We present the results of IRS low-resolution spectroscopy of 51 Seyfert
galaxies, part of a large \emph{Spitzer} observing program to determine the
mid-to-far infrared spectral energy distributions of a well-defined sample of
87 nearby, 12~\micron-selected Seyferts.  We find that the spectra clearly
divide into groups based on their continuum shapes and spectral features.  The
infrared spectral types appear to be related to the Seyfert types. Some
features are clearly related to a starburst contribution to the IR spectrum,
while the observed power-law continuum shapes, attributed to the AGN, may be
dust or non-thermal emission.  Principal component analysis results suggest
that the relative contribution of starburst emission is the dominant cause
of variance in the spectra.  We find that the Sy 2's  show on average
stronger starburst contributions than the Sy 1's.  
\end{abstract}

\section{Introduction}

Understanding the inner workings of Active Galactic Nuclei (AGNs) is
important for understanding the role of supermassive black holes in
galaxy evolution. The mid-to-far infrared (MFIR) is a key wavelength
regime for testing the AGN unified scheme and probing the nature of
the nuclear obscuring medium. The \emph{Spitzer Space Telescope}
allows studies to be made at higher sensitivity and spatial resolution
that previous IR missions\footnote{This work is based on observations
made with the Spitzer Space Telescope, which is operated by the Jet
Propulsion Laboratory, California Institute of Technology under a
contract with NASA.}.  We are conducting a large observing program
with all three \emph{Spitzer} instruments to determine the MFIR
spectral energy distributions (SEDs) of 87 Seyfert galaxies (PI:
J. Gallimore, PID: 3269).  The sample of Seyfert galaxies used for
this study comprises all Seyfert galaxies from the extended
12~\micron\ sample of \citet{rus93} that have $cz <
10000$~km\,s$^{-1}$.  Here we discuss the 51 objects in the sample for
which the low-resolution, 5 -- 35~\micron\ IRS spectra are currently
available\footnote{The IRS was a collaborative venture between Cornell
University and Ball Aerospace Corporation funded by NASA through the
Jet Propulsion Laboratory and Ames Research Center.}. The analysis of
these spectra and a more detailed discussion of the results are
presented in \citet{buc06}.

\section{Infared Spectral Shapes and Seyfert Types}

\begin{figure}
\plotfiddle{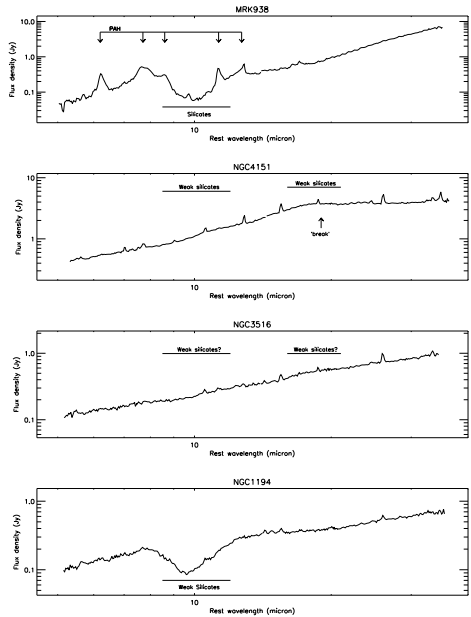}{3.1 in}{0}{30.0}{30.0}{-100}{0}
\caption{Typical spectra in each group: Mrk 938 (red continuum with PAH
  features), NGC4151 (broken power-law), NGC3516 (power-law), and NGC1194
  (silicate absorption). All spectra have been converted to restframe.
  \label{fig:egspec}} 
\end{figure}

We find four distinct types of continuum shapes and spectral features
among the IRS spectra of the Seyfert nuclei.  Figure \ref{fig:egspec}
shows typical spectra in each group.  We group the spectra according
to the following main properties:\\
$\bullet$ Twenty-four objects (47\%\ of the observed objects) have
spectra dominated by strong PAH emission features and very red
continuum suggestive of cool dust; Mrk\,938 is the archetype.  Many
(16) of these objects show clear or weak silicate absorption at
10~\micron.  \\
$\bullet$ Sixteen objects (31\%) have continua that can be described
by a broken power-law; they show a flattening in the continuum slope
at $\sim$20~\micron; NGC\,4151 is the brightest of this class.  This
flattening may be due to a warm ($\sim$170~K) dust component, peaking
at $\sim$20~\micron, dominating the mid-IR emission (e.g.,
\citealt{wee05,per01,rod96}), however, a simple model of an
emissivity-weighted blackbody function does not fit the continuum of
these spectra well.  Two of these objects, Mrk6 and Mrk335, show clear
silicate emission features. A further 9 objects in this group,
including NGC4151, show weak excesses at 10 and 18~\micron\ that may
be due to silicate emission. \\
$\bullet$ Eight objects (16\%) have power-law continuous spectra.
NGC~3516 is representative of this class. These spectra show no strong
dust emission or absorption features, though most have weak excesses
at 10 and 18~\micron\ that may be due to silicate emission. \\
$\bullet$ Two objects (NGC\,1194 \& F04385$-$0828) show a broad
absorption feature at 9.7~\micron\ due to silicate dust.  \\
One object (NGC\,7603) appears to show both PAH and silicate emission
features and otherwise fails to meet any of the above classifications.

\begin{figure}
\protect\vspace*{-0.8 in}
\plottwo{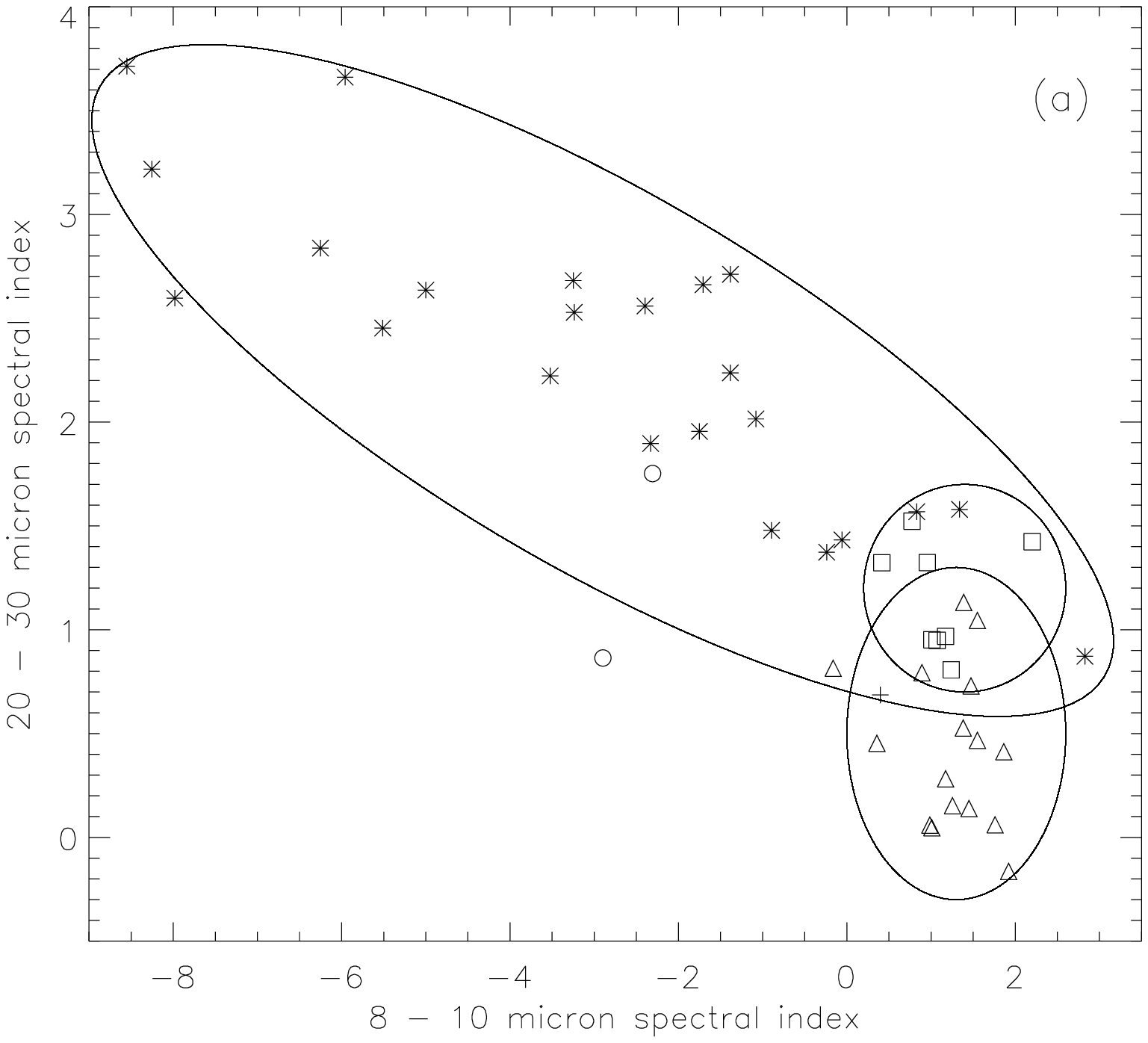}{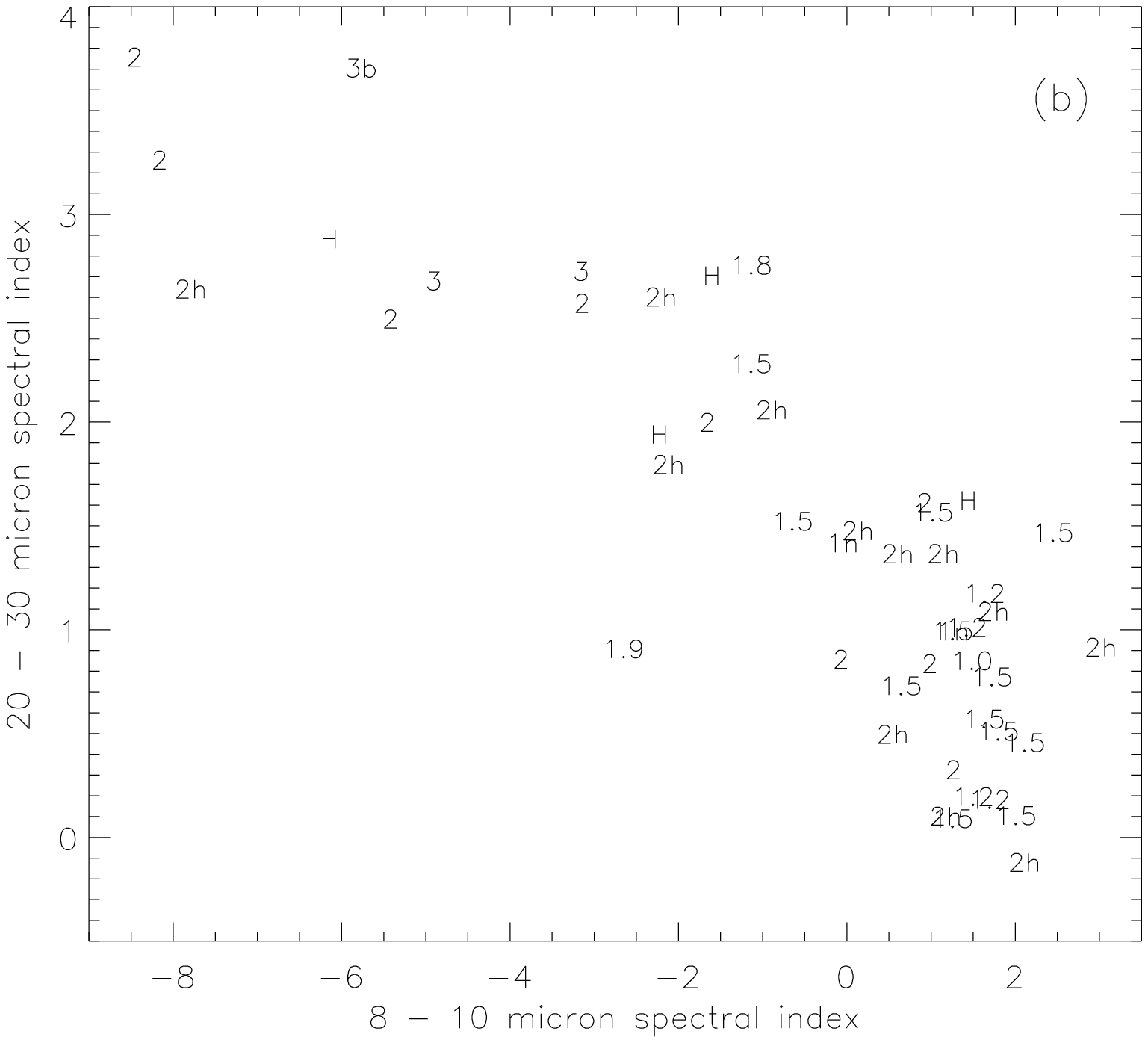}
\protect\vspace*{-0.5 in}
\caption{IR color-color diagrams derived from the IRS spectra. {\it
  (a)} The symbols indicate the shape of the spectrum: red continuum
  with PAH features {\it (stars)}, broken power-law {\it (triangles)},
  unbroken power-law {\it (squares)}, and silicate absorption {\it
  (circles)}. The ellipses show the areas occupied by the three
  largest groups. {\it (b)} The numbers indicate the Seyfert spectral
  type. Numbers 1.0 -- 1.9 and 1n represent Seyfert 1
  subtypes. Symbols 2h and 2 represent Seyfert 2's with and without
  HBLRs, respectively. LINERs, with and without broad lines, are shown
  by numbers 3 and 3b, respectively, and starburst galaxies are
  indicated by the letter H.  \label{fig:cols}}
\end{figure}

The spectral shapes may be quantitatively characterized by the IR colors, or
spectral indices, derived from the spectra. The spectral index between 20 and
30~\micron\ quantifies the slope of the spectrum (above the $\sim$20~\micron\
`break' for broken power-law spectra), while the spectral index between 8 and
10~\micron\ indicates the presence or absence of the 9.7~\micron\ silicate
feature and PAH features. The resulting color-color diagrams are given in
Figure \ref{fig:cols}.  One Sy 2 (NGC3079) with extreme colors,
$\alpha_{20-30\,\micron} = 4.7$ and $\alpha_{8-10\,\micron} = -12.5$, is not
shown.  The Sy 1's (Seyfert types 1.0, 1.2 and 1.5) lie in the lower-right
part of the diagram, occupied by unbroken and broken power-law spectra,
specifically avoiding the region of the reddest PAH emitters. Sy 2's are found
in both regions. Earlier type Seyferts (i.e. 1 -- 1.5) have the bluest colors,
while Seyfert 2's with hidden broad line regions (HBLRs; type 2h) span the
range of colors.  The fact that Sy 1's consistently show the bluest infrared
colors and the Sy 2's are the reddest sources indicates the IR spectra are
changing systematically with Seyfert type.

\begin{figure}
\plotfiddle{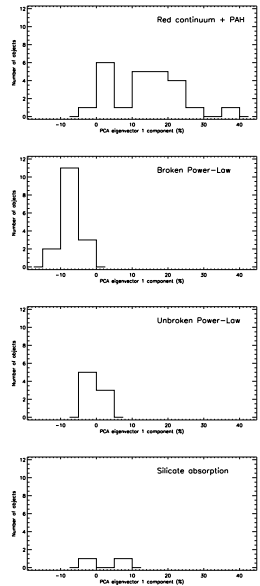}{2.7 in}{0}{35.0}{35.0}{-40}{0}
\caption{Distribution of the relative contribution of eigenvector 1
  for the different shapes of the mid-infrared SED in the sample.
  \label{fig:evonehist}}
\end{figure}

We performed a principal component analysis (PCA) on the spectra to
determine the component spectral shapes (eigenvectors) producing the
variety of spectra seen in the sample (see, e.g., \citealt{fra99,
sha04}). We find that the first eigenvector is the most dominant and
contributes 91\% of the variance in the spectra.   We
find that the relative contribution of eigenvector 1 to each spectrum
relates strongly to the shape of the SED, as the different shapes have
significantly different contributions from this eigenvector (Figure
\ref{fig:evonehist}).

\begin{figure}
\plotfiddle{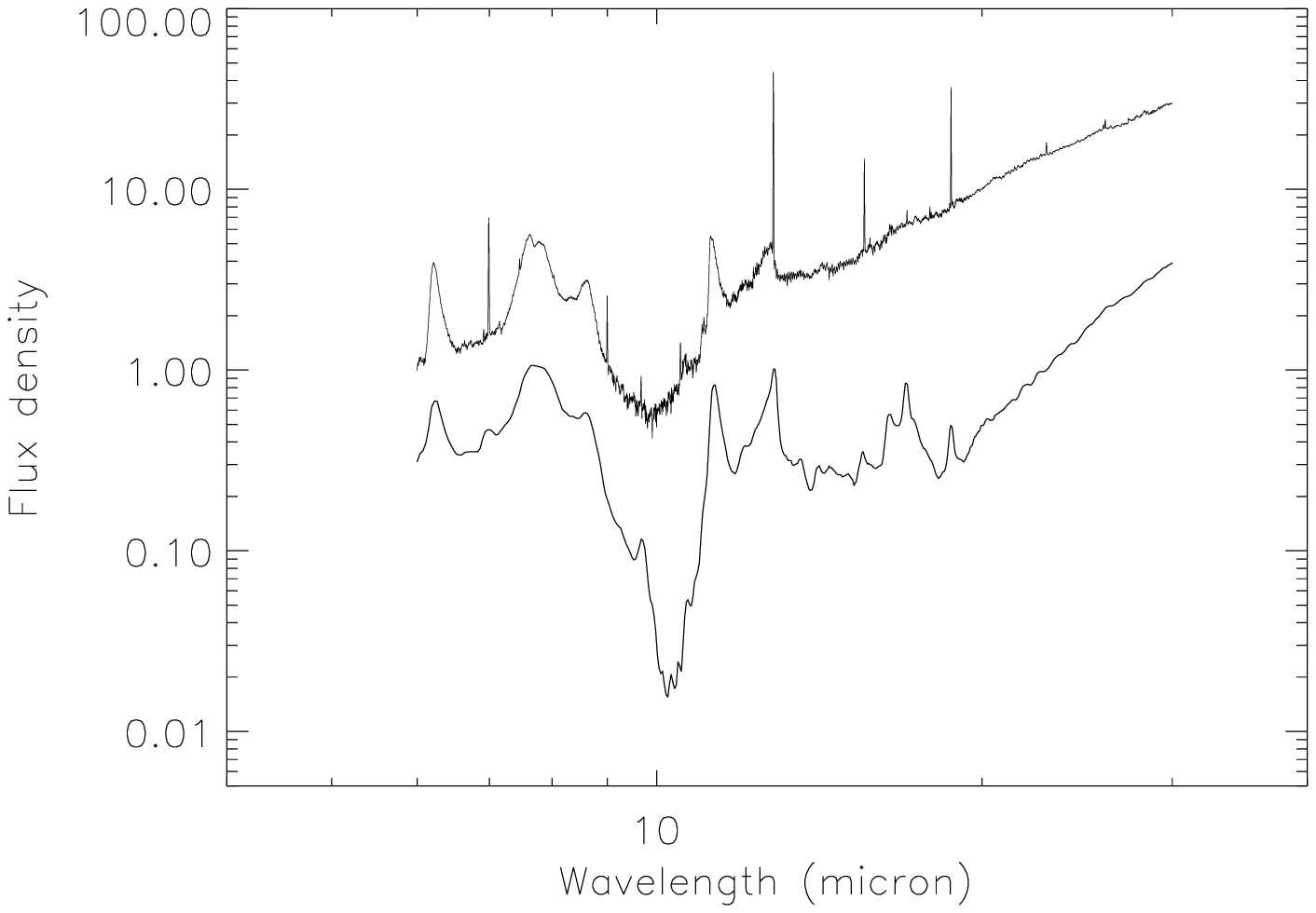}{1.8 in}{0}{40.0}{40.0}{-125}{0}
\protect\vspace*{-10mm}
\caption{The first eigenvector of the PCA which we interpret
  as dominated by a starburst component but also including
  contribution from other features {\it (thin line)}, and, for
  comparison, the ISO spectrum of starburst galaxy M82 {\it (thick
  line)}, from
  http://isc.astro.cornell.edu/~sloan/library/swsatlas/atlas.html. The
  eigenvector clearly shows the PAH features and silicate absorption
  seen in starburst galaxies. The shape of the eigenvector at the
  longer wavelength differs slightly from a starburst shape. It is
  possible that this shape in the eigenvector produces the break in
  slope in the NGC~4151-like objects.  \label{fig:sbfigs}}
\end{figure}

\begin{figure}
\plotfiddle{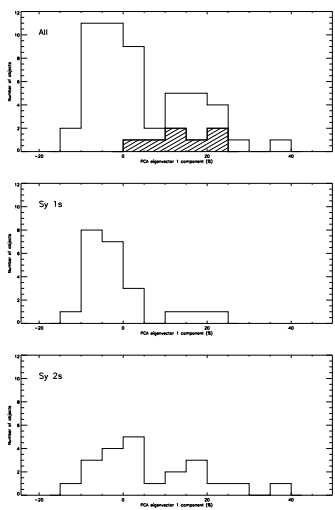}{2.5 in}{0}{30.0}{30.0}{-60}{0}
\caption{Distribution of the relative contribution of eigenvector 1 for all
  observed objects {\it (top)}, Sy 1's {\it (middle)} and Sy 2's {\it
  (bottom)}. The hashed region in the top panel indicates the objects
  reclassified as non-Seyferts (LINERs and starburst galaxies).
  \label{fig:evonesy}} 
\end{figure}

The similarity in the features between the red objects like MRK~938
and the IR spectra of starburst galaxies such as M~82 strongly
suggests these objects are dominated by the starburst contribution to
the dust heating at mid-infrared wavelengths (Figure
\ref{fig:sbfigs}).    We
find a significant difference between the contribution of the first
eigenvector to the spectra of the Sy 1's and Sy 2's (Figure
\ref{fig:evonesy}).  Our results confirm the previous findings of
\citet{mai95} and \citet{gor04}, based on ground-based photometry,
that Sy 2's show more star formation than Sy 1's.

\section{Conclusions}

We present the results of Spitzer IRS nuclear spectra for the first 51 of a
complete sample of 12~\micron\ selected Seyfert galaxies.  We find the
following main conclusions: \\
$\bullet$ The spectra clearly divide into groups based on their
continuum shapes and spectral features. The largest group (47\% of the
sample of 51) shows very red continuum suggestive of cool dust and
strong emission features attributed to PAHs. Sixteen objects (31\%)
are well described by a broken power-law continuum. A further 16\% of
the sample show power-law continua with unchanging slopes.  Two
objects are dominated by a broad silicate absorption feature. One
object in the sample shows an unusual spectrum dominated by emission
features, that is unlike any of the other spectra. Some features are
clearly related to the starburst contribution to the IR spectrum,
while the mechanisms producing the power-law continuum attributed to
the AGN component are not yet clear. The IR SEDs are related to the
Seyfert types but the precise correspondence between the IR types and
the optical spectral types is unclear. \\
$\bullet$ Principal component analysis suggests that the relative
contribution of starburst-heated dust emission to the SED is the
dominant cause of variance in the observed spectra. \\
$\bullet$ We find that the Sy 2's typically show stronger starburst
contributions in their IR spectra than the Sy 1's, confirming previous
results found using photometric data.  \\
Detailed modeling of the continuum emission is underway to separate in
detail the starburst and AGN contributions to the IR spectrum in order
to place constraints on the opacity and geometry of the nuclear
obscuring material, and to compare the relative starburst
contributions of Seyfert types 1 and 2.

\acknowledgements 

We thank the conference organizers for an excellent meeting.  Support
for this work was provided by NASA through an award issued by
JPL/Caltech.

\protect


\vspace*{-2 mm}
\begin{thebibliography}{}
\bibitem[Antonucci(1993)]{ant93} Antonucci, R.\ 1993, \araa, 31, 473
\bibitem[Buchanan et al.(2006)]{buc06} Buchanan, C.\ L., Gallimore, J.\ F.,
  O'Dea, C.\ P., Baum, S.\ A., Axon, D.\ J., Robinson, A., Elitzur, M., \&
  Elvis, M.\ 2006, \aj, submitted
\bibitem[Francis \& Wills(1999)]{fra99} Francis, P.~J., \& Wills, B.~J.\ 1999,
ASP Conf.~Ser.~162: Quasars and Cosmology, 162, 363
\bibitem[Gorjian et al.(2004)]{gor04} Gorjian, et al. 2004, \apj, 605, 156
\bibitem[Maiolino et al.(1995)]{mai95} Maiolino, R., Ruiz, M., Rieke,
G.\ H., Keller, L.\ D.\ 1995, \apj, 446, 561
\bibitem[Perez Garcia \& Rodriguez Espinosa(2001)]{per01} Perez
  Garcia, A.\ M., Rodriguez Espinosa, J.\ M.\ 2001, \apj, 557, 39
\bibitem[Rodriguez Espinosa et al.(1996)]{rod96} Rodriguez Espinosa,
J.\ M., Perez Garcia, A.\ M., Lemke, D., Meisenheimer, K.\ 1996, \aap,
315, L129
\bibitem[Rush et al.(1993)]{rus93} Rush, B., Malkan, M.~A., \& Spinoglio, L.\
1993, \apjs, 89, 1
\bibitem[Shang \& Wills(2004)]{sha04} Shang, Z., \& Wills, B.\ 2004, ASP
Conf.~Ser.~311: AGN Physics with the Sloan Digital Sky Survey, 311, 13
\bibitem[Urry \& Padovani(1995)]{urr95} Urry, C.~M., \& Padovani, P.\ 1995,
\pasp, 107, 803
\bibitem[Weedman et al.(2005)]{wee05} Weedman, D.~W., et al.\ 2005, \apj, 633,
706
\end{thebibliography}
\end{document}